\documentclass[aps,onecolumn,10pt]{revtex4}
\usepackage{epsfig}
\usepackage{graphicx}
\usepackage{amsmath}
\usepackage{amssymb}
\usepackage{hyperref}
\hypersetup{colorlinks=true,linkcolor=red,citecolor=green,urlcolor=blue}
\numberwithin{equation}{section}
\numberwithin{equation}{section}

\begin{document}
\allowdisplaybreaks
\setcounter{equation}{0}

\title{Normalization of the vacuum and the ultraviolet completion of Einstein gravity}

\author{Philip D. Mannheim}
\affiliation{Department of Physics, University of Connecticut, Storrs, CT 06269, USA \\
philip.mannheim@uconn.edu\\ }

\date{March 19 2023}

\begin{abstract}
Second-order-derivative plus fourth-order-derivative gravity is the ultraviolet completion of second-order-derivative quantum Einstein gravity. While it achieves renormalizability through states of negative Dirac norm, the unitarity violation  that this would entail can be postponed to Planck energies. As we show in this paper the theory has a different problem, one that occurs at all energy scales, namely that the Dirac norm of the vacuum of the theory is not finite. To establish this we present a procedure for determining the norm of the vacuum in any quantum field theory. With the Dirac norm of the vacuum of the second-order-derivative plus fourth-order-derivative theory not being finite, the Feynman rules that are used to establish renormalizability are not valid, as is the assumption that the theory can be used as an effective theory at energies well below the Planck scale. This lack of finiteness is also manifested in the fact that the  Minkowski path integral for the theory is divergent. Because the vacuum Dirac norm is not finite, the Hamiltonian of the theory is not Hermitian. However, it turns out to be $PT$ symmetric. And when one continues the theory into the complex plane and uses the $PT$ symmetry inner product, viz.  the overlap of the left-eigenstate of the Hamiltonian with its right-eigenstate, one then finds that for the vacuum this norm is both finite and positive, the Feynman rules now are valid, the Minkowski path integral now is well behaved, and the theory now can serve as a low energy effective theory. Consequently, the theory can now be offered as a fully consistent, unitary and renormalizable theory of quantum gravity.

\end{abstract}

\maketitle

Essay written for the Gravity Research Foundation 2023 Awards for Essays on Gravitation.

\section{The hidden assumption of quantum field theory}
\label{S1}

For a quantum field theory to be viable it must be formulated in a Hilbert space whose inner product is time independent, finite, and  positive (though zero norm is also acceptable). While any inner product that meets these three requirements would be satisfactory, the most common choice that is made is the Dirac one, viz. the overlap of a ket with its Hermitian conjugate bra. While it is usually straightforward to check for the sign and time dependence of  the Dirac norm in any given theory, it is not straightforward to check for its finiteness, and often finiteness is just taken for granted. However, for the consistency of the theory, this is something that has to be checked, and taking the Dirac norm to be finite is actually a hidden assumption. In this paper we shall present a procedure for making such a check. We find that the Dirac norm of second-order-derivative, neutral scalar quantum field theory is finite, but that of a second-order-derivative plus fourth-order-derivative, neutral scalar field theory, a theory that shares many of the features of  quantum gravity, is not. Then we discuss how to resolve this concern.

To illustrate the issues involved consider a free, relativistic, second-order-derivative, neutral scalar quantum field theory with action, wave equation, Hamiltonian, and equal time commutation relation of the form  
\begin{align}
&I_{\rm S}=\displaystyle{\int}d^4x \tfrac{1}{2}\left[\partial_{\mu}\phi\partial^{\mu}\phi-m^2\phi^2\right],\qquad[\partial_{\mu}\partial^{\mu}+m^2]\phi=0,
\nonumber\\
&H=\displaystyle{\int } d^3x \tfrac{1}{2}[\dot{\phi}^2+\bar{\nabla}\phi\cdot \bar{\nabla}\phi+m^2\phi^2],\qquad [\phi(\bar{x},t),\dot{\phi}(\bar{x}^{\prime},t)]=i\delta^3(\bar{x}-\bar{x}^{\prime}).
\label{1.2}
\end{align}
With $\omega_k=+(\bar{k}^2+m^2)^{1/2}$ solutions to the  wave equation  obey 
\begin{eqnarray}
\phi(\bar{x},t)=\int \frac{d^3k}{\sqrt{(2\pi)^3 2\omega_k}}[a(\bar{k})e^{-i\omega_k t+i\bar{k}\cdot\bar{x}}+a^{\dagger}(\bar{k})e^{i\omega_k  t-i\bar{k}\cdot\bar{x}}],
\label{1.3}
\end{eqnarray}
 and with $[a(\bar{k}),a^{\dagger}(\bar{k}^{\prime})]=\delta^3(\bar{k}-\bar{k}^{\prime})$ the Hamiltonian is given by 
 \begin{eqnarray}
 H=\frac{1}{2}\int d^3k[\bar{k}^2+m^2]^{1/2} \left[a^{\dagger}(\bar{k})a(\bar{k})+a(\bar{k})a^{\dagger}(\bar{k})\right].
 \label{1.4}
 \end{eqnarray}
Given (\ref{1.4}) we can introduce a no-particle state $\vert \Omega\rangle$ that obeys $a(\bar{k})\vert \Omega \rangle=0$ for each $\bar{k}$, and can identify it as the ground state of $H$. This procedure does not specify the value of $\langle \Omega\vert \Omega\rangle$.

For the theory the associated c-number propagator obeys
\begin{align}
&(\partial_t^2-\bar{\nabla}^2+m^2)D(x)=-\delta^4(x),
\label{1.5}
\end{align}
so that
\begin{align}
& D(x)
=\int \frac{d^4k}{(2\pi)^4}\frac{e^{-ik\cdot x}}{(k^2-m^2+i\epsilon)}.
\label{1.6}
\end{align}
If we identify the  propagator as a vacuum matrix element of q-number fields, viz. 
\begin{eqnarray}
D(x)=-i\langle \Omega\vert T[\phi(x)\phi(0)]\vert\Omega\rangle,
\label{1.7}
\end{eqnarray}
then use of the equal time commutation relation gives 
\begin{align}
&(\partial_t^2-\bar{\nabla}^2+m^2)(-i)\langle \Omega\vert T[\phi(x)\phi(0)]\vert\Omega\rangle=-\langle \Omega \vert\Omega \rangle \delta^4(x). 
\label{1.8}
\end{align}
Comparing with (\ref{1.5}) we see that we can only identify $D(x)$ as the matrix element $-i\langle \Omega\vert T[\phi(x)\phi(0)]\vert\Omega\rangle$ if the vacuum is normalized to one, viz. $\langle \Omega\vert \Omega\rangle=1$. Now if the normalization of the vacuum is finite we of course can always rescale it to one. However, that presupposes that the normalization of the vacuum  is not infinite. We are not aware of any proof in the literature that the Dirac norm of the vacuum is not infinite (either in this particular case or in general), and taking it to be finite is a hidden assumption. So in this paper, which is based on  \cite{Mannheim2022,Mannheim2023}, we shall present a procedure for determining whether the normalization of the vacuum state is or is not finite, a procedure that is built not on quantum field theory but on quantum mechanics.

\section{The quantum-mechanical simple harmonic oscillator} 
\label{S2}

For a quantum-mechanical simple harmonic oscillator with Hamiltonian $H=\tfrac{1}{2}[p^2+q^2]$ and commutator $[q,p]=i$, there are two sets of bases, the wave function basis and the occupation number space basis. The wave function basis is obtained by setting $p=-i\partial/\partial q$ in $H$ and then solving  the Schr\"odinger wave equation $H\psi(q)=E\psi(q)$. In this way we obtain a ground state with energy $E_0=\tfrac{1}{2}$ and wave function $\psi_0(q)=e^{-q^2/2}/\pi^{1/4}$. For occupation number space we set $q=(a+a^{\dagger})/\sqrt{2}$ and $p=i(a^{\dagger}-a)/\sqrt{2}$. This yields $[a,a^{\dagger}]=1$ and $H=a^{\dagger} a+1/2$. We introduce a no-particle state $\vert \Omega \rangle$ that obeys $a\vert \Omega \rangle =0$, with $\vert \Omega \rangle$ being the occupation number space ground state with energy $E_0=\tfrac{1}{2}$.  However, in and of itself this does not fix the norm $\langle \Omega\vert \Omega\rangle$ of the no-particle state or oblige it to be finite.

To fix the $\langle \Omega\vert \Omega\rangle$ norm we need to relate the ground states of the two bases. With $a=(q+ip)/\sqrt{2}$ we set
\begin{align}
\langle q \vert a\vert \Omega \rangle= \frac{1}{\sqrt{2}}\left(q+\frac{\partial}{\partial q}\right)\langle q \vert \Omega \rangle=0,
\label{2.1}
\end{align}
and find that $ \langle q \vert \Omega \rangle=e^{-q^2/2}$. We thus identify $\psi_0(q)=\langle q \vert \Omega \rangle$. We now calculate the standard Dirac norm for the vacuum, and obtain  
\begin{align}
\langle \Omega \vert \Omega \rangle=\int_{-\infty}^{\infty}dq \langle \Omega \vert q\rangle\langle q\vert \Omega \rangle=
\int_{-\infty}^{\infty}dq \psi^*_0(q)\psi_0(q)
=\int_{-\infty}^{\infty} dq e^{-q^2}=\pi^{1/2}.
\label{2.2}
\end{align}
We thus establish that the Dirac norm of the no-particle state is finite. And on setting $\psi_0(q)=e^{-q^2/2}/\pi^{1/4}$ we normalize it to one. That we are able to do this is because we know the form of the wave function $\psi_0(q)$.  

While this  procedure is both straightforward and familiar, it works because both the wave function basis approach and occupation number basis approach have something in common, namely that they are both based on an infinite number of degrees of freedom. For the occupation number basis we can represent the creation and annihilation operators as infinite-dimensional matrices labeled by $\vert \Omega\rangle$, $a^{\dagger}\vert \Omega\rangle$, $a^{\dagger 2}\vert \Omega\rangle$ and so on. For the wave function basis the coordinate $q$ is a continuous variable that varies between $-\infty$ and $\infty$. The two sets of bases are both infinite dimensional, one discrete and the other continuous. The advantage of the continuous basis is that it enables us to express the normalization of the vacuum state as an integral with an infinite range, an integral that is then either finite or infinite.

\section{The quantum field theory oscillator} 
\label{S3}

In the quantum field theory case we do not know the form of the wave function solutions to $H\vert \psi\rangle=E\vert \psi\rangle$, since we cannot realize the canonical commutator given in (\ref{1.2}) as a differential relation. Specifically, we cannot satisfy (\ref{1.2}) by setting $\dot{\phi}(\bar{x},t)$ equal to $-i\partial /\partial \phi(\bar{x},t)$ (though we could introduce a functional derivative $\dot{\phi}(\bar{x},t)=-i\delta /\delta \phi(\bar{x},t)$).

However, we can express the Hamiltonian in terms of creation and annihilation operators.   
So what we can then do is reverse engineer what  we did in the quantum-mechanical case.  We thus introduce 
\begin{align}
&a(\bar{k})= \frac{1}{\sqrt{2}}[q(\bar{k})+ip(\bar{k})],\quad a^{\dagger}(\bar{k})= \frac{1}{\sqrt{2}}[q(\bar{k})-ip(\bar{k})],
\nonumber\\
&[q(\bar{k}),p(\bar{k}^{\prime})]=i\delta^3(\bar{k}-\bar{k}^{\prime}),\quad H=\frac{1}{2}\int d^3k[\bar{k}^2+m^2]^{1/2} [p^2(\bar{k})+q^2(\bar{k})],
\nonumber\\
&\phi(\bar{x},t)=\frac{1}{\sqrt{2}}\int \frac{d^3k}{\sqrt{(2\pi)^3 2\omega_k}}\left[[q(\bar{k})+ip(\bar{k})]e^{-i\omega_k t+i\bar{k}\cdot\bar{x}}+[q(\bar{k})-ip(\bar{k})]e^{i\omega_k  t-i\bar{k}\cdot\bar{x}}\right].
\label{3.2}
\end{align}
These $q(\bar{k})$ and $p(\bar{k})$ operators bear no relation to any physical position or momentum operators. Their only role here is to enable us to convert the discrete infinite-dimensional basis associated with each $a(\bar{k})$ and $a^{\dagger}(\bar{k})$ into a continuous one. Specifically, we can realize the $[q(\bar{k}),p(\bar{k}^{\prime})]$ commutator by $p(\bar{k}^{\prime})=-i\partial/\partial q(\bar{k}^{\prime})$, with $H$ then becoming a wave operator. In this way  for each $\bar{k}$ we obtain a solution to the Schr\"odinger equation of the form $\psi(\bar{k})=e^{-q^2(\bar{k})/2}/\pi^{1/4}$. We can define a no-particle vacuum that obeys $a(\bar{k})\vert \Omega\rangle$ for each $\bar{k}$. For each $\bar{k}$ we have
\begin{align}
\langle q(\bar{k})\vert a(\bar{k})\vert \Omega \rangle= \frac{1}{\sqrt{2}}\left[q(\bar{k})+\frac{\partial}{\partial q(\bar{k})}\right]\langle q(\bar{k})\vert \Omega \rangle=0,
\label{3.3}
\end{align}
so that $\langle q(\bar{k})\vert \Omega\rangle=e^{-q^2(\bar{k})/2}/\pi^{1/4}$, and thus
\begin{align}
\langle \Omega\vert \Omega \rangle=\Pi_{\bar{k}}\int d q(\bar{k})\langle \Omega\vert q(\bar{k}) \rangle\langle q(\bar{k})\vert \Omega \rangle
=\Pi_{\bar{k}}\int d q(\bar{k})\frac{e^{-q^2(\bar{k})}}{\pi^{1/2}}=\Pi_{\bar{k}}1=1.
\label{3.4}
\end{align}
Thus the vacuum for the full $H$ obeys $\langle \Omega\vert\Omega\rangle=1$, to thus have a finite normalization. In this way we establish that the vacuum state of the free, relativistic, second-order-derivative, neutral scalar field theory is normalizable.

The general prescription then is to convert the occupation number space Hamiltonian into a product of individual occupation number spaces each with its own $\bar{k}$, and then determine whether the equivalent wave mechanics ground state wave functions constructed this way have a finite normalization in the conventional Schr\"odinger wave mechanics theory sense. If they do, then so does the full  vacuum $\vert \Omega \rangle$ of the full $H$. If on the other hand the equivalent wave mechanics wave functions are not normalizable, then neither is the full $\vert \Omega \rangle$.  

Once we are able to show that the vacuum state of the free theory is normalizable, this will remain true in the presence of interactions if the interacting theory is renormalizable. To see this we note that  in developing Wick's contraction theorem in quantum field theory one needs to put the time-ordered product of Heisenberg fields $\phi(x)$, viz. $
\tau(x_1,...,x_n)=\langle \Omega \vert T[\phi(x_1)...\phi(x_n)]\vert \Omega \rangle$, into a form that can be developed perturbatively. To this end one introduces a set of in-fields $\phi_{in}(x)$ that satisfy free field equations with Hamiltonian $H_{in}$. And one also introduces an evolution operator $U(t)$ that evolves the interaction Hamiltonian $H_I(t)$ and fields according to
\begin{align}
i\frac{\partial U(t)}{\partial t}=H_I(t)U(t),\qquad
\phi(\bar{x},t)=U^{-1}(t)\phi_{in}(\bar{x},t)U(t).
\label{9.3}
\end{align}
Using these relations we obtain (see e.g. \cite{Bjorken1965}) 
\begin{align}
\tau(x_1,...,x_n)&=\langle \Omega \vert T\left[\phi_{in}(x_1)...\phi_{in}(x_n)\exp\left(-i\int_{-t}^tdt_1H_I(t_1)\right)\right]\vert \Omega \rangle
\langle \Omega \vert T\left[\exp\left(i\int_{-t}^tdt_1H_I(t_1)\right)\right]\vert \Omega \rangle.
\label{9.6}
\end{align}
After inverting the last term we obtain the standard form for the perturbative Wick contraction procedure, viz. 
\begin{align}
\tau(x_1,...,x_n)&=\frac{\langle \Omega \vert T\left[\phi_{in}(x_1)...\phi_{in}(x_n)\exp\left(-i\int_{-t}^tdt_1H_I(t_1)\right)\right]\vert \Omega \rangle}
{\langle \Omega \vert T\left[\exp\left(-i\int_{-t}^tdt_1H_I(t_1)\right)\right]\vert \Omega \rangle}.
\label{9.7}
\end{align}
If one starts with (\ref{9.7}) it would appear that the normalization of the vacuum state is actually irrelevant since it would drop out of the ratio. And so it would not appear to matter if it did happen to be infinite. However, this is not the case since we could only go from (\ref{9.6}) to (\ref{9.7}) if $\langle \Omega \vert T\left[\exp\left(i\int_{-t}^tdt_1H_I(t_1)\right)\right]\vert \Omega \rangle$ is finite. And it would not be if the vacuum state is not normalizable. If we expand $\langle \Omega \vert T\left[\exp\left(i\int_{-t}^tdt_1H_I(t_1)\right)\right]\vert \Omega \rangle$ out as a power series in $H_I$ the first term is $\langle \Omega \vert  \Omega \rangle$ as calculated in a free theory. Thus,  for finiteness we first need this term to be finite and then need the power series expansion in $H_I$ to be renormalizable in order for the interacting $\langle \Omega \vert T\left[\exp\left(i\int_{-t}^tdt_1H_I(t_1)\right)\right]\vert \Omega \rangle$ to be finite too.  However, for a nonnormalizable vacuum the standard Wick expansion and Feynman rules that are obtained from (\ref{9.7}) are not valid. Since this concern is of relevance to radiative corrections to Einstein gravity we return to this point in Sec. \ref{S11}.

As well as providing a procedure for determining whether or not $\langle \Omega\vert \Omega\rangle$ is finite, since the procedure enables is to express the free second-order-derivative Hamiltonian $H$ as an ordinary derivative operator, it does so for interactions as well. Specifically, from (\ref{3.2}) we can write $\phi(\bar{x},t)$ as a derivative operator, viz.
\begin{align}
&\phi(\bar{x},t)=\frac{1}{\sqrt{2}}\int \frac{d^3k}{\sqrt{(2\pi)^3 2\omega_k}}\left[\left[q(\bar{k})+\frac{\partial}{\partial q(\bar{k})}\right]e^{-i\omega_k t+i\bar{k}\cdot\bar{x}}+\left[q(\bar{k})-\frac{\partial}{\partial q(\bar{k})}\right]e^{i\omega_k  t-i\bar{k}\cdot\bar{x}}\right].
\label{3.5}
\end{align}
Thus the insertion of (\ref{3.5}) into an interaction Hamiltonian of the form $H_I=\lambda \int d^3x \phi^4(\bar{x},t)$ enables us to write $H_I$, and thus $H+H_I$,  as a derivative operator. While this procedure enables us to in principle set up the Schr\"odinger problem for $H+H_I$ as a wave mechanics problem, it is still quite a formidable one, just as interacting field theories always have been.

\section{Higher-derivative quantum field theories}
\label{S4}

Having presented an example of a theory whose vacuum Dirac norm is finite, we now present an example for which $\langle \Omega\vert \Omega\rangle$ is not finite. The example is based on a second-order-derivative plus fourth-order-derivative, neutral scalar field theory with action and equation of motion 
\begin{align}
I_S&=\frac{1}{2}\int d^4x\bigg{[}\partial_{\mu}\partial_{\nu}\phi\partial^{\mu}
\partial^{\nu}\phi-(M_1^2+M_2^2)\partial_{\mu}\phi\partial^{\mu}\phi
+M_1^2M_2^2\phi^2\bigg{]},\quad (\partial_t^2-\bar{\nabla}^2+M_1^2)(\partial_t^2-\bar{\nabla}^2+M_2^2)
\phi(x)=0,
\label{4.1}
\end{align}
with ${\rm diag}[\eta_{\mu\nu}]=(1,-1,-1,-1)$.  While we now study this particular model just for illustrative purposes, we note that  it actually arises in quantum gravity studies, and in Sec. \ref{S11} we shall explore the implications of this study for quantum gravity. For (\ref{4.1}) the associated propagator obeys
\begin{align}
&(\partial_t^2-\bar{\nabla}^2+M_1^2)(\partial_t^2-\bar{\nabla}^2+M_2^2)D(x)=-\delta^4(x),
\nonumber\\
&D(x)=-\int \frac{d^4k}{(2\pi)^4}\frac{e^{-ik\cdot x}}{(k^2-M_1^2)(k^2-M_2^2)}
=-\int \frac{d^4k}{(2\pi)^4}\frac{e^{-ik\cdot x}}{(M_1^2-M_2^2)}\left[\frac{1}{(k^2-M_1^2)}-\frac{1}{(k^2-M_2^2)}\right].
\label{4.2}
\end{align}
The energy-momentum tensor $T_{\mu\nu}$, the canonical momenta $\pi^{\mu}$ and $\pi^{\mu\lambda}$, and the equal-time commutators appropriate to the higher-derivative theory are given by \cite{Bender2008b} 
\begin{align}
T_{\mu\nu}&=\pi_{\mu}\phi_{,\nu}+\pi_{\mu}^{~\lambda}\phi_{,\nu,\lambda}-\eta_{\mu\nu}{\cal L},
\nonumber\\ 
 \pi^{\mu}&=\frac{\partial{\cal L}}{\partial \phi_{,\mu}}-\partial_{\lambda
}\left(\frac{\partial {\cal L}}{\partial\phi_{,\mu,\lambda}}\right)=-\partial_{\lambda}\partial^{\mu}\partial^{\lambda}\phi- (M_1^2+M_2^2)\partial^{\mu}\phi,\qquad  \pi^{\mu\lambda}=\frac{\partial {\cal L}}{\partial \phi_{,\mu,\lambda}}=\partial^{\mu}\partial^{\lambda}\phi,
\nonumber\\
T_{00}&=\tfrac{1}{2}\pi_{00}^2+\pi_{0}\dot{\phi}+\tfrac{1}{2}(M_1^2+M_2^2)\dot{
\phi}^2-\tfrac{1}{2}M_1^2M_2^2\phi^2
-\tfrac{1}{2}\pi_{ij}\pi^{ij}+\tfrac{1}{2}(M_1^2+M_2^2)\phi_{,i}\phi^{,i}
\nonumber\\
&=\frac{1}{2}\ddot{\phi}^2-\tfrac{1}{2}(M_1^2+M_2^2)\dot{
\phi}^2-\dddot{\phi}\dot{\phi}-[\partial_i\partial^i\dot{\phi}]\dot{\phi}
-\tfrac{1}{2}M_1^2M_2^2\phi^2
-\tfrac{1}{2}\partial_i\partial_j\phi\partial^i\partial^j\phi+\tfrac{1}{2}(M_1^2+M_2^2)\partial_i\phi\partial^i\phi,
\nonumber\\
&[\phi(\bar{0},t),\dot{\phi}(\bar{x},t)]=0, \qquad[\phi(\bar{0},t),\ddot{\phi}(\bar{x},t)]=0, \qquad [\phi(\bar{0},t),\dddot{\phi}(\bar{x},t])=-i\delta^3(x).
\label{4.3}
\end{align}
With the use of these commutation relations we find that $D(x)=i\langle \Omega\vert T[\phi(x)\phi(0)]\vert\Omega\rangle$
indeed satisfies the first equation given in (\ref{4.2}), provided that is  that $\langle \Omega\vert \Omega\rangle=1$ \cite{footnote1}.

To check whether  $\langle \Omega\vert \Omega\rangle$ actually is finite, we need to express the scalar field Hamiltonian $H_S=\int d^3x T_{00}$ in terms of creation and annihilation operators and then construct an equivalent wave mechanics. Given that the solutions to (\ref{4.1}) are plane waves,  we set
\begin{eqnarray}
\phi(\bar{x},t)=\int \frac{d^3k}{(2\pi)^{3/2}}\left[a_1(\bar{k})e^{-i\omega_1 t+i\bar{k}\cdot\bar{x}}+a^{\dagger}_1(\bar{k})e^{i\omega_1  t-i\bar{k}\cdot\bar{x}}+a_2(\bar{k})e^{-i\omega_2 t+i\bar{k}\cdot\bar{x}}+a^{\dagger}_2(\bar{k})e^{i\omega_2  t-i\bar{k}\cdot\bar{x}}\right],
\label{4.5}
\end{eqnarray}
where $\omega_1=+(\bar{k}^2+M_1^2)^{1/2}$, $\omega_2=+(\bar{k}^2+M_2^2)^{1/2}$. Given (\ref{4.5}) and the commutators in (\ref{4.3}) we obtain 
\begin{align}
H_S&=(M_1^2-M_2^2)\int d^3k\bigg{[}(\bar{k}^2+M_1^2)\left[a^{\dagger}_{1}(\bar{k})a_1(\bar{k})
+a_{1}(\bar{k})a^{\dagger}_1(\bar{k})\right]
-(\bar{k}^2+M_2^2)\left[a^{\dagger}_2(\bar{k})a_{2}(\bar{k})
+a_{2}(\bar{k})a^{\dagger}_2(\bar{k})\right]\bigg{]},
\nonumber\\
&[a_1(\bar{k}),a^{\dagger}_{1}(\bar{k}^{\prime})]=[2(M_1^2-M_2^2)(\bar{k}^2+
M_1^2)^{1/2}]^{-1}\delta^3(\bar{k}-\bar{k}^{\prime}),
\nonumber\\
&[a_2(\bar{k}),a^{\dagger}_{2}(\bar{k}^{\prime})]=-[2(M_1^2-M_2^2)(\bar{k}^2+
M_2^2)^{1/2}]^{-1}\delta^3(\bar{k}-\bar{k}^{\prime}),
\nonumber\\
&[a_1(\bar{k}),a_{2}(\bar{k}^{\prime})]=0,\quad[a_1(\bar{k}),a^{\dagger}_{2}(\bar{k}^{\prime})]=0,\quad[a^{\dagger}_1(\bar{k}),a_{2}(\bar{k}^{\prime})]=0,\quad
[a^{\dagger}_1(\bar{k}),a^{\dagger}_{2}(\bar{k}^{\prime})]=0.
\label{4.6}
\end{align}
We note that with $M_1^2-M_2^2>0$ for definitiveness, we see negative signs in both $H_S$ and the $[a_2(\bar{k}),a^{\dagger}_{2}(\bar{k}^{\prime})]$ commutator.  We shall see below that the negative sign concerns will be resolved once we settle the issue of the normalization of the vacuum. To do that we descend to the quantum-mechanical limit of the theory, the Pais-Uhlenbeck oscillator model.

\section{Higher-derivative quantum mechanics} 
\label{S5}

In order to study the Pauli-Villars regulator, in \cite{Pais1950} Pais and Uhlenbeck (${\rm PU}$) introduced a fourth-order quantum-mechanical oscillator model with action and equation of motion
\begin{eqnarray}
I_{\rm PU}=\frac{1}{2}\int dt\left[{\ddot z}^2-\left(\omega_1^2
+\omega_2^2\right){\dot z}^2+\omega_1^2\omega_2^2z^2\right],\qquad \ddddot{z}+(\omega_1^2+\omega_2^2)\ddot{z}+\omega_1\omega_2z^2=0,
\label{5.1}
\end{eqnarray}
where for definitiveness in the following we take $\omega_1>\omega_2$. As constructed this action possesses three variables $z$, $\dot{z}$ and $\ddot{z}$. This is too many for one oscillator but not enough for two. The system is thus a constrained system. And so we introduce a new variable $x=\dot{z}$ and its conjugate $p_x$. And using the method of Dirac constraints for Poisson brackets, following canonical quantization we obtain the time-independent quantum Hamiltonian and canonical equal-time commutators  \cite{Mannheim2000,Mannheim2005}
\begin{align}
H_{\rm PU}&=\frac{p_x^2(t)}{2}+p_z(t)x(t)+\frac{1}{2}\left(\omega_1^2+\omega_2^2 \right)x^2(t)-\frac{1}{2}\omega_1^2\omega_2^2z^2(t),\qquad 
[z(t),p_z(t)]=i, \qquad [x(t),p_x(t)]=i.
\label{5.3}
\end{align}
The terms in $H_{\rm PU}$ are in complete parallel to the first four terms in the field theory $T_{00}$ given in (\ref{4.3}), with the PU oscillator model being the nonrelativistic limit of the relativistic scalar field theory, with the spatial dependence having been frozen out. Since canonical commutators only involve time derivatives, freezing out the spatial dependence will still give the full dynamical content of the relativistic theory. In fact we can set $i=[z,p_z]\equiv [\phi,\pi_0]=[\phi,-\dddot{\phi}-(M_1^2+M_2^2)\dot{\phi}]=i\delta^3(x)$, to thus parallel the commutators given in (\ref{4.3}).    

On setting $p_z=-i\partial_z$, $p_x=-i\partial_x$ the Schr\"odinger problem for $H_{\rm PU}$ can be solved analytically, with
the state with energy $(\omega_1+\omega_2)/2$ having a wave function that is of the form \cite{Mannheim2007} 
\begin{align}
\psi_0(z,x)=\exp[\tfrac{1}{2}(\omega_1+\omega_2)\omega_1\omega_2z^2+i\omega_1\omega_2zx-\tfrac{1}{2}(\omega_1+\omega_2)x^2].
\label{5.4}
\end{align}
While this wave function is well behaved at large $x$, it diverges at large $z$, and consequently  it  is not normalizable. 

To relate this wave function to the no-particle vacuum $\vert \Omega \rangle$ we second quantize the theory. And with the wave equation given in (\ref{5.1}), and with $\dot{z}=i[H_{\rm PU},z]=x$, $\dot{x}=p_x$, $\dot{p}_x=-p_z-(\omega_1^2+\omega_2^2)x$, $\dot{p}_z=\omega_1^2\omega_2^2z$, we obtain 
\begin{align}
z(t)&=a_1e^{-i\omega_1t}+a_1^{\dagger}e^{i\omega_1t}+a_2e^{-i\omega_2t}+a_2^{\dagger}e^{i\omega_2t},\quad
p_z(t)=i\omega_1\omega_2^2
[a_1e^{-i\omega_1t}-a_1^{\dagger}e^{i\omega_1t}]+i\omega_1^2\omega_2[a_2e^{-i\omega_2t}-a_2^{\dagger}e^{i\omega_2t}],
\nonumber\\
x(t)&=-i\omega_1[a_1e^{-i\omega_1t}-a_1^{\dagger}e^{i\omega_1t}]-i\omega_2[a_2e^{-i\omega_2t}-a_2^{\dagger}e^{i\omega_2t}],\quad
p_x(t)=-\omega_1^2 [a_1e^{-i\omega_1t}+a_1^{\dagger}e^{i\omega_1t}]-\omega_2^2[a_2e^{-i\omega_2t}+a_2^{\dagger}e^{i\omega_2t}],
\label{5.5}
\end{align}
and a Hamiltonian and commutator algebra of the form \cite{Mannheim2000,Mannheim2005}
\begin{align}
H_{\rm PU}&=2(\omega_1^2-\omega_2^2)(\omega_1^2 a_1^{\dagger}
a_1-\omega_2^2a_2^{\dagger} a_2)
+\tfrac{1}{2}(\omega_1+\omega_2),
\nonumber\\
[a_1,a_1^{\dagger}]&=\frac{1}{2\omega_1(\omega_1^2-\omega_2^2)}, \qquad
[a_2,a_2^{\dagger}]=-\frac{1}{2\omega_2(\omega_1^2-\omega_2^2)}.
\label{5.7}
\end{align}
We note the similarity to (\ref{4.6}). 

In the Hilbert space in which both $a_1$ and $a_2$ annihilate the vacuum the energy spectrum is bounded from below, and the energy of the ground state is $(\omega_1+\omega_2)/2$ with wave function $\psi_0(z,x)e^{-i(\omega_1+\omega_2)t/2}$, where $\psi_0(z,x)$is given in (\ref{5.4}). For this wave function the normalization of $\vert \Omega \rangle$ is then  given by 
\begin{align}
\langle \Omega\vert \Omega\rangle=\int_{-\infty}^{\infty} dz\int_{-\infty}^{\infty}dx\langle \Omega\vert z,x\rangle\langle z,x\vert\Omega\rangle=\int_{-\infty}^{\infty} dz \int_{-\infty}^{\infty}dx \psi_0^*(z,x)\psi_0(z,x).
\label{5.8}
\end{align}
With $\psi_0(z,x)$ diverging at large $z$, this normalization integral is infinite. Thus we see that through our knowledge of the form of the ground state wave function as given in (\ref{5.4}) we are able to determine the normalization of the PU theory vacuum and  establish that it is infinite. We can thus anticipate and will immediately show in Sec. \ref{S6} that this is also the case for the second-order plus fourth-order scalar quantum field  theory as well. Then in Sec. \ref{S7} we will discuss what to do about it, with there actually being a mechanism for obtaining a finite normalization \cite{Bender2008a,Bender2008b}, one that also takes care of the fact that according to (\ref{5.7}) $\langle \Omega\vert a_2a_2^{\dagger}\vert\Omega\rangle$ is negative.

\section{The nonnormalizable vacuum of higher-derivative field theories}
\label{S6}
 To determine the second-order plus fourth-order scalar field theory vacuum normalization we introduce
\begin{align}
z(\bar{k},t)&=a_1(\bar{k})e^{-i\omega_1(\bar{k})t}+a_1^{\dagger}(\bar{k})e^{i\omega_1(\bar{k})t}+a_2(\bar{k})e^{-i\omega_2(\bar{k})t}+a_2^{\dagger}(\bar{k})^{i\omega_2(\bar{k})t},
\nonumber\\
p_z(\bar{k},t)&=i\omega_1(\bar{k})\omega_2^2(\bar{k})
[a_1(\bar{k})e^{-i\omega_1(\bar{k})t}-a_1^{\dagger}(\bar{k})e^{i\omega_1(\bar{k})t}]+i\omega_1^2(\bar{k})\omega_2(\bar{k})[a_2(\bar{k})e^{-i\omega_2(\bar{k})t}-a_2^{\dagger}(\bar{k})e^{i\omega_2(\bar{k})t}],
\nonumber\\
x(\bar{k},t)&=-i\omega_1(\bar{k})[a_1(\bar{k})e^{-i\omega_1(\bar{k})t}-a_1^{\dagger}(\bar{k})e^{i\omega_1(\bar{k})t}]-i\omega_2(\bar{k})[a_2(\bar{k})e^{-i\omega_2(\bar{k})t}-a_2^{\dagger}(\bar{k})^{i\omega_2(\bar{k})t}],
\nonumber\\
p_x(\bar{k},t)&=-\omega_1^2(\bar{k})[a_1(\bar{k})e^{-i\omega_1(\bar{k})t}+a_1^{\dagger}(\bar{k})e^{i\omega_1(\bar{k})t}]-\omega_2^2(\bar{k})[a_2(\bar{k})e^{-i\omega_2(\bar{k})t}+a_2^{\dagger}(\bar{k})^{i\omega_2(\bar{k})t}].
\label{6.3}
\end{align}
From (\ref{6.3}) and the commutation relations given in (\ref{4.6}) it follows that 
\begin{eqnarray}
&&[z(\bar{k},t),p_z(\bar{k}^{\prime},t)]=i\delta^3(\bar{k}-\bar{k}^{\prime}),\qquad [x(\bar{k},t),p_x(\bar{k}^{\prime},t)]=i\delta^3(\bar{k}-\bar{k}^{\prime}),
\nonumber\\
&&[z(\bar{k},t),x(\bar{k}^{\prime},t)]=0,\quad[z(\bar{k},t),p_x(\bar{k}^{\prime},t)]=0,\quad[p_z(\bar{k},t),x(\bar{k}^{\prime},t)]=0,\quad
[p_z(\bar{k},t),p_x(\bar{k}^{\prime},t)]=0.
\label{6.4}
\end{eqnarray}
Given (\ref{6.3}) we can rewrite the Hamiltonian given in (\ref{4.6}) in the equivalent, time-independent form
\begin{eqnarray}
H_S&=&\int d^3k\bigg{[}\frac{p_x^2(\bar{k},t)}{2}+p_z(\bar{k},t)x(\bar{k},t)+\frac{1}{2}\left[\omega_1^2(\bar{k})+\omega_2^2(\bar{k}) \right]x^2(\bar{k},t)-\frac{1}{2}\omega_1^2(\bar{k})\omega_2^2(\bar{k})z^2(\bar{k},t)\bigg{]}.
\label{6.5}
\end{eqnarray}
For each momentum state we recognize the quantum field theory Hamiltonian $H_S$ given in (\ref{6.5}) as being of precisely the form of the quantum-mechanical $H_{\rm PU}$ Hamiltonian that is given in (\ref{5.3}).

We can now proceed as in the second-order scalar  theory discussed above and represent the commutators by
\begin{eqnarray}
&&\left[z(\bar{k},t), -i\frac{\partial}{\partial z(\bar{k}^{\prime},t)}\right]=\delta^3(\bar{k}-\bar{k}^{\prime}),\qquad \left[x(\bar{k},t),-i\frac{\partial}{\partial x(\bar{k}^{\prime},t)}\right]=\delta^3(\bar{k}-\bar{k}^{\prime}).
\label{6.6}
\end{eqnarray}
With the vacuum obeying $a_1(\bar{k})\vert \Omega\rangle=0$, $a_2(\bar{k})\vert \Omega\rangle=0$ for each $\bar{k}$, from (\ref{6.3})  we obtain 
\begin{align}
&\langle z(\bar{k}),x(\bar{k})\vert a_1(\bar{k})\vert \Omega\rangle=\frac{1}{2(M_1^2-M_2^2)}\left[-\omega_2^2(\bar{k})z(\bar{k})+i\frac{\partial}{\partial x(\bar{k})}+i\omega_1(\bar{k})x(\bar{k})+\frac{1}{\omega_1(\bar{k})}\frac{\partial}{\partial z(\bar{k})}\right]\langle z(\bar{k}),x(\bar{k})\vert\Omega\rangle=0,
\nonumber\\
&\langle z(\bar{k}),x(\bar{k})\vert a_2(\bar{k})\vert \Omega\rangle=\frac{1}{2(M_1^2-M_2^2)}\left[\omega_1^2(\bar{k})z(\bar{k})-i\frac{\partial}{\partial x(\bar{k})}-i\omega_2(\bar{k})x(\bar{k})-\frac{1}{\omega_2(\bar{k})}\frac{\partial}{\partial z(\bar{k})}\right]\langle z(\bar{k}),x(\bar{k})\vert\Omega\rangle=0,
\label{6.7}
\end{align}
for each $\bar{k}$. From (\ref{6.7}) it follows that for each $\bar{k}$ we can identify each $\langle z(\bar{k}),x(\bar{k})\vert\Omega\rangle$ with the PU oscillator ground state wave function $\psi_0(z(\bar{k}),x(\bar{k}))$, which, analogously to (\ref{5.4}), is given by
\begin{align}
\psi_0(z(\bar{k}),x(\bar{k}))=\exp[\tfrac{1}{2}[\omega_1(\bar{k})+\omega_2(\bar{k})]\omega_1(\bar{k})\omega_2(\bar{k})z^2(\bar{k})+i\omega_1(\bar{k})\omega_2(\bar{k})z(\bar{k})x(\bar{k})-\tfrac{1}{2}[\omega_1(\bar{k})+\omega_2(\bar{k})]x^2(\bar{k})].
\label{6.8}
\end{align}
Consequently,  the Dirac norm of the vacuum is given by
\begin{align}
\langle \Omega\vert \Omega\rangle&=\Pi_{\bar{k}}\int_{-\infty}^{\infty} dz(\bar{k})\int_{-\infty}^{\infty}dx(\bar{k})\langle \Omega\vert  z(\bar{k}),x(\bar{k})\rangle\langle z(\bar{k}),x(\bar{k})\vert\Omega\rangle
\nonumber\\
&=\Pi_{\bar{k}}\int_{-\infty}^{\infty} dz(\bar{k}) \int_{-\infty}^{\infty}dx(\bar{k}) \psi_0^*(z(\bar{k}),x(\bar{k}))\psi_0(z(\bar{k}),x(\bar{k})).
\label{6.9}
\end{align}
With each $\psi_0(z(\bar{k}),x(\bar{k}))$ diverging at large $z(\bar{k})$, we thus establish that the Dirac norm of the field theory vacuum is infinite. Thus whatever is the normalization of the vacuum in the associated wave-mechanical limit translates into the same normalization in the quantum field theory.

\section{How to obtain a normalizable vacuum}
\label{S7}

Rather than herald the doom of the second-order plus fourth-order theory, we note that the above analysis has a shortcoming in it, namely the assumption of the Hermiticity and self-adjointness of the operators in the theory.  Specifically, to show that an operator is self-adjoint we need to be able to throw away surface terms in an integration by parts. This we cannot do if wave functions are not normalizable. The $H_S$ Hamiltonian is thus not Hermitian. If  $H_S$ is not Hermitian then for any arbitrary state $\vert \psi(t)\rangle$ in the Hilbert space we have $\langle \psi(t)\vert\psi(t)\rangle= \langle \psi(0)\vert e^{iH_S^{\dagger}t}e^{-iH_St}\psi(0)\rangle$, which is not equal to $\langle \psi(0)\vert \psi(0)\rangle$. Consequently, for the non-Hermitian $H_S$ the Dirac inner product is not time independent, and is thus not appropriate. For the theory to be viable we have to find some other inner product that is time independent, and at the same time need it to be both finite and nonnegative.

For the field $\phi(\bar{x},t)$ we note from (\ref{4.6}) that the matrix element  $\langle \Omega \vert a_2(\bar{k})a_2^{\dagger}((\bar{k})\vert \Omega \rangle$ is negative. This is not possible if $a_2^{\dagger}(\bar{k})$ is the Hermitian conjugate of $a_2(\bar{k})$ since the matrix element would then have to be positive definite. It is thus invalid to identify the coefficient of the $e^{i\omega_2t-i\bar{k}\cdot\bar{x}}$ term in the expansion of $\phi(\bar{x},t)$ as given in (\ref{4.5}) as the Hermitian conjugate of the coefficient of the $e^{-i\omega_2t+i\bar{k}\cdot\bar{x}}$ term. Consequently the field $\phi(\bar{x},t)$ cannot be Hermitian, and thus the $H_S$ that is constructed from it cannot be Hermitian either.

To see this difficulty in an alternate way  consider the Minkowski time path integral associated with the field theory action given in (\ref{4.1}),  viz.
\begin{align}
PI(MINK)=\int D[\phi]D[\sigma_{\mu}]\exp\left[\frac{i}{2}\int_{-\infty}^{\infty} d^4x\left[\partial_{\nu}\sigma_{\mu}\partial^{\nu}\sigma^{\mu}-\left(M_1^2+M_2^2\right)\sigma_{\mu}\sigma^{\mu}+M_1^2M_2^2\phi^2\right]\right],
\label{12.5}
\end{align}
where $\sigma_{\mu}=\partial_{\mu}\phi$. (In the same way that we treat $z$ and $x=\dot{z}$ as independent degrees of freedom in the PU case we must do the same thing for $\phi$ and $\sigma_{\mu}=\partial_{\mu}\phi$.)

In order to damp out oscillations and enable the path integral to be well defined we need to introduce an $i\epsilon$ prescription. To determine an appropriate one we note that  if we replace $M_1^2$ and $M_2^2$ by $M_1^2-i\epsilon$, $M_2^2-i\epsilon$ in the $1/(k^2-M_1^2)-1/(k^2-M_2^2)$ propagator given in (\ref{4.2}),  then positive frequency poles in the complex $k^0$ plane will propagate forward in time while negative frequency poles will propagate backward (i.e., energies bounded from below). However the residues in the $M_2$ sector will be negative (the negative norm problem). Alternatively, if we replace  $M_1^2$ and $M_2^2$ by $M_1^2-i\epsilon$, $M_2^2+i\epsilon$ in the $1/(k^2-M_1^2)-1/(k^2-M_2^2)$ propagator,  then $M_2$ sector positive frequency poles in the complex $k^0$  plane will propagate backward in time while negative frequency poles will propagate forward (i.e., energies unbounded from below, the Ostrogradski instability of higher derivative theories). However then the  $M_2$ sector poles will be traversed in a way that would make the residues be positive. Inspection of (\ref{4.6})  indicates that these two realizations respectively correspond to separate and distinct Hilbert spaces in which $a_2(\bar{k})$ or $a^{\dagger}_2(\bar{k})$ annihilates the vacuum. While we thus have both negative energy and negative residue problems, in no Hilbert space do we have both.

For  the $M_1^2-i\epsilon$, $M_2^2-i\epsilon$ prescription the path integral takes the form
\begin{align}
PI(MINK)&=\int D[\phi]D[\sigma_{\mu}]\exp\bigg{[}\frac{1}{2}\int_{-\infty}^{\infty} d^4x\big{[}i\partial_{\nu}\sigma_{\mu}\partial^{\nu}\sigma^{\mu}-i\left(M_1^2+M_2^2\right)\sigma_{\mu}\sigma^{\mu}+iM_1^2M_2^2\phi^2
\nonumber\\
&-2\epsilon\sigma_{\mu}\sigma^{\mu}+(M_1^2+M_2^2)\epsilon\phi^2\big{]}\bigg{]}.
\label{12.6}
\end{align}
With $\phi$ and $\sigma_{\mu}$ being taken to be real and with $\sigma_{\mu}\sigma^{\mu}$ being taken to be timelike on every path, the $\sigma_{\mu}$ path integration is damped but  the $\phi$ path integration is not. 

For the unconventional $i\epsilon$ prescription in which we replace $M_1^2$ and $M_2^2$ by $M_1^2-i\epsilon$, $M_2^2+i\epsilon$ the path integral takes the form
\begin{align}
PI(MINK)&=\int D[\phi]D[\sigma_{\mu}]\exp\bigg{[}\frac{1}{2}\int_{-\infty}^{\infty} d^4x\big{[}i\partial_{\nu}\sigma_{\mu}\partial^{\nu}\sigma^{\mu}-i\left(M_1^2+M_2^2\right)\sigma_{\mu}\sigma^{\mu}+iM_1^2M_2^2\phi^2-(M_1^2-M_2^2)\epsilon\phi^2\big{]}\bigg{]},
\label{12.7}
\end{align}
and has no damping on the $\sigma_{\mu}$ path integration at all. This unconventional $i\epsilon$ prescription for the Feynman contour leads to an unbounded from below energy spectrum that cannot be associated with a well-defined path integral at all, and so we do not consider it any further.

Thus the only Feynman $i\epsilon$ prescription that can be relevant is the standard one with $M_1^2-i\epsilon$, $M_2^2-i\epsilon$. However, even with this choice the $\phi$ path integration is not damped if $\phi$ is real. To rectify this we need find an appropriate domain in the complex plane for the measure in which the $\phi$ path integration is damped. And in fact the path integral  does become damped if we do not require $\phi$ to be real, but instead take it to be pure imaginary \cite{Bender2008b,mannheim2018antilinearity} (though $({\rm Im}[\phi])^2> ({\rm Re}[\phi])^2$ would suffice). With the $\phi\rightarrow -i\phi= \bar{\phi}$ transformation being achieved by a complex classical symplectic transformation that preserves the classical Poisson bracket algebra, we replace (\ref{12.6}) by
\begin{align}
PI(MINK)&=\int D[\bar{\phi}]D[\sigma_{\mu}]\exp\bigg{[}\frac{1}{2}\int_{-\infty}^{\infty} d^4x\big{[}i\partial_{\nu}\sigma_{\mu}\partial^{\nu}\sigma^{\mu}-i\left(M_1^2+M_2^2\right)\sigma_{\mu}\sigma^{\mu}-iM_1^2M_2^2\bar{\phi}^2
\nonumber\\
&-2\epsilon\sigma_{\mu}\sigma^{\mu}-(M_1^2+M_2^2)\epsilon\bar{\phi}^2\big{]}\bigg{]}.
\label{12.8}
\end{align}
With $\bar{\phi}$ and $\sigma_{\mu}$ being taken to be real and with $\sigma_{\mu}\sigma^{\mu}$ being taken to be timelike on every path, the path integral is now well defined, and can now be associated with a well-defined quantum theory \cite{footnote2a}.

However, we still need to determine what particular quantum theory the now damped path integral might correspond to. To this end we note that the wave function $\psi_0(z(\bar{k}),x(\bar{k}))$ given in (\ref{6.8}) would become normalizable if we replace $z(\bar{k})$ by $-iz(\bar{k})$, while making no change in $x(\bar{k})$. To achieve this change we make a quantum similarity transformation (viz. the quantum counterpart of a classical symplectic transformation) on the quantum fields of the form \cite{Bender2008a,mannheim2018antilinearity}
$S(S)=e^{\pi\int d^3x\pi_0(\bar{x},t)\phi(\bar{x},t)/2}$. Such a transformation preserves both energy eigenvalues and canonical commutators, and effects
\begin{align}
S(S)\phi(\bar{x},t)S(S)^{-1}=-i\phi(\bar{x},t)=\bar{\phi}(\bar{x},t),\qquad S(S)\pi(\bar{x},t)S(S)^{-1}=i\pi(\bar{x},t)=\bar{\pi}(\bar{x},t),
\nonumber\\
S(S)z(\bar{k})S(S)^{-1}=-iz(\bar{k})\equiv y(\bar{k}),\qquad S(S)p_z(\bar{k})S(S)^{-1}=ip_z(\bar{k})\equiv q(\bar{k}).
\label{7.2}
\end{align}

Now, even though $H$ and also $\bar{H}_S$ (see (\ref{7.15}) below) are not Hermitian, all of their eigenvalues are real, as must be the case since all the poles of  the $1/(k^2-M_1^2)-1/(k^2-M_2^2)$ propagator are on the real $k^0$ axis, and their locations do not change under a similarity transformation. Now common as its use is,  a Hermiticity condition is only sufficient to secure real eigenvalues but not necessary. (While Hermitian Hamiltonians have real eigenvalues, there is no converse theorem that says that a non-Hermitian Hamiltonian must have at least one  complex eigenvalue.) However, there is a necessary condition for real eigenvalues, namely that  \cite{bender2010PT,mannheim2018antilinearity} the Hamiltonian have an antilinear symmetry  \cite{footnote2}. The second-order-derivative plus fourth-order-derivative theory thus falls into the class of $PT$  theories ($P$ is the linear parity operator and $T$ is the antilinear time reversal operator) developed by Bender and collaborators \cite{bender2007making,bender2019pt}. (In general, even without Hermiticity one still has $CPT$ symmetry \cite{mannheim2018antilinearity}, but since the scalar fields are neutral $C$ is separately conserved, so in this case $CPT$ defaults to $PT$.) Critical to the $PT$ program  is that the wave functions be normalizable in some domain in the complex plane, a domain known technically as a Stokes wedge. Since the wave functions are not normalizable with real $z(\bar{k})$, we have to continue  $z(\bar{k})$ into the complex plane in order to make them normalizable. Then, precisely just as we found with the path integral measure,  the theory is then well defined. Thus $\bar{H}_S$ must be $PT$ symmetric, and $\bar{\phi}(\bar{x},t)$ will be $PT$ odd ($PT\phi(\bar{x},t)TP=-\bar{\phi}(-\bar{x},-t)$) since the starting $\phi(\bar{x},t)$ is a standard $P$ even, $T$ even neutral scalar field.

To see how $PT$ symmetry is explicitly implemented we  note that the wave equation given in (\ref{4.1}) does not change under the $\phi(\bar{x},t)\rightarrow -i\bar{\phi}(\bar{x},t)$ transformation. We can thus expand $\bar{\phi}(\bar{x},t)$ in a complete basis of plane waves as 
\begin{align}
\bar{\phi}(x)&=\int \frac{d^3k}{(2\pi)^{3/2}}\left [-ia_1(\bar{k})e^{-i\omega_1(\bar{k})t+i\bar{k}\cdot \bar{x}}+a_2(\bar{k})e^{-i\omega_2(\bar{k}) t+i\bar{k}\cdot \bar{x}}-i\hat{a}_1(\bar{k})e^{i\omega_1(\bar{k}) t-i\bar{k}\cdot \bar{x}}+\hat{a}_2(\bar{k})e^{i\omega_2(\bar{k}) t-i\bar{k}\cdot \bar{x}}\right].
\label{7.11}
\end{align}
 In (\ref{7.11}) we have introduced a hatted notation to indicate that the hatted operators are not Hermitian conjugates of the unhatted ones. As constructed, $\bar{\phi}(\bar{x},t)$ will be $PT$ odd if the creation and annihilation operators obey 
\begin{align}
PTa_1(\bar{k})TP=a_1(\bar{k}),\quad PT\hat{a}_1(\bar{k})TP=\hat{a}_1(\bar{k}), \quad PTa_2(\bar{k})TP=-a_2(\bar{k}),\quad PT\hat{a}_2(\bar{k})TP=-\hat{a}_2(\bar{k}).
\label{7.11a}
\end{align}
Given (\ref{7.11}),  the transformed Hamiltonian $\bar{H}_S=S(S)H_SS(S)^{-1}$ and transformed commutators now take the form \cite{Bender2008b}
\begin{align}\bar{H}_S&=(M_1^2-M_2^2)\int d^3k\bigg{[}(\bar{k}^2+M_1^2)
\left[\hat{a}_{1}(\bar{k})a_1(\bar{k})+a_{1}(\bar{k})\hat{a}_1(\bar{k})\right]
+(\bar{k}^2+M_2^2)\left[\hat{a}_{2}(\bar{k})a_2(\bar{k})+a_{2}(\bar{k})\hat{a}_2(\bar{k})\right]\bigg{]},
\nonumber\\
& [\dot{\bar{\phi}}(\bar{x},t),\bar{\phi}(0)]=0,\qquad [\ddot{\bar{\phi}}(\bar{x},t),\bar{\phi}(0)]=0,\qquad [\dddot{\bar{\phi}}(\bar{x},t),\bar{\phi}(0)]=i\delta^3(x),
\nonumber\\
&[a_1(\bar{k}),\hat{a}_{1}(\bar{k}^{\prime})]=[2(M_1^2-M_2^2)(\bar{k}^2+
M_1^2)^{1/2}]^{-1}\delta^3(\bar{k}-\bar{k}^{\prime}),
\nonumber\\
&[a_2(\bar{k}),\hat{a}_{2}(\bar{k}^{\prime})]=[2(M_1^2-M_2^2)(\bar{k}^2+
M_2^2)^{1/2}]^{-1}\delta^3(\bar{k}-\bar{k}^{\prime}),
\nonumber\\
&[a_1(\bar{k}),a_{2}(\bar{k}^{\prime})]=0,
\quad [a_1(\bar{k}),\hat{a}_{2}(\bar{k}^{\prime})]=0,
\quad [\hat{a}_{1}(\bar{k}),a_{2}(\bar{k}^{\prime})]=0,
\quad [\hat{a}_{1}(\bar{k}),\hat{a}_{2}(\bar{k}^{\prime})]=0,
\label{7.13}
\end{align}
and with (\ref{7.11a}) it follows that all of these relations are $PT$ symmetric. The algebra of the creation and annihilation operators as given in (\ref{7.13}) thus provides a faithful representation of the field commutation relations.
With all relative signs in (\ref{7.13}) being positive (we take $M_1^2>M_2^2$ for definitiveness), there are no states of negative norm or of negative energy, and the theory is now fully viable. As such, this discussion completely parallels the previously published discussion of the $PU$ oscillator model given in \cite{Bender2008a,Bender2008b}.

To complete the field theory discussion we replace (\ref{6.3}), (\ref{6.4}) and (\ref{6.5}) by 
\begin{align}
y(\bar{k},t)&=-ia_1(\bar{k})e^{-i\omega_1(\bar{k})t}+a_2(\bar{k})e^{-i\omega_2(\bar{k})t}-i\hat{a}_1(\bar{k})e^{i\omega_1(\bar{k})t}+\hat{a}_2(\bar{k})
e^{i\omega_2(\bar{k})t},
\nonumber\\
x(\bar{k},t)&=-i\omega_1(\bar{k})a_1(\bar{k})e^{-i\omega_1(\bar{k})t}+\omega_2(\bar{k})a_2(\bar{k})e^{-i\omega_2(\bar{k})t}+i\omega_1(\bar{k})\hat{a}_1(\bar{k})
e^{i\omega_1(\bar{k})t}-\omega_2(\bar{k})\hat{a}_2(\bar{k})e^{i\omega_2(\bar{k})t},
\nonumber\\
p_x(\bar{k},t)&=-\omega_1^2(\bar{k})a_1(\bar{k})e^{-i\omega_1(\bar{k})t}-i\omega_2^2(\bar{k})a_2(\bar{k})e^{-i\omega_2(\bar{k})t}-\omega_1
^2(\bar{k})\hat{a}_1(\bar{k})e^{i\omega_1(\bar{k})t}-i\omega_2^2(\bar{k})\hat{a}_2(\bar{k})e^{i\omega_2(\bar{k})t},
\nonumber\\
q(\bar{k},t)&=\omega_1(\bar{k})\omega_2(\bar{k})[-\omega_2(\bar{k})a_1(\bar{k})e^{-i\omega_1(\bar{k})t}-i\omega_1(\bar{k})a_2(\bar{k})e^{-i\omega_2(\bar{k})t}+\omega_2(\bar{k})\hat{a}_1(\bar{k})e^{i\omega_1(\bar{k})t}+i\omega_1(\bar{k})\hat{a}_2(\bar{k})e^{i\omega_2(\bar{k})t}],
\nonumber\\
&[y(\bar{k},t),q(\bar{k}^{\prime},t)]=i\delta^3(\bar{k}-\bar{k}^{\prime}),\qquad [x(\bar{k},t),p_x(\bar{k}^{\prime},t)]=i\delta^3(\bar{k}-\bar{k}^{\prime}),
\nonumber\\
\bar{H}_S&=\int d^3k\bigg{[}\frac{p_x^2(\bar{k},t)}{2}-iq(\bar{k},t)x(\bar{k},t)+\frac{1}{2}\left[\omega_1^2(\bar{k})+\omega_2^2(\bar{k}) \right]x^2(\bar{k},t)+\frac{1}{2}\omega_1^2(\bar{k})\omega_2^2(\bar{k})y^2(\bar{k},t)\bigg{]}.
\label{7.15}
\end{align}
In (\ref{7.15}) $PT$ symmetry is maintained by having $y(\bar{k},t)$  and $x(\bar{k},t)$ be $PT$ odd, and $p_x(\bar{k},t)$  and $q(\bar{k},t)$ be $PT$ even, while the factor of $i$ in (\ref{7.15}) shows that $\bar{H}_S$ is not Hermitian \cite{footnote5a}. Compared with (\ref{6.3}) we note that only two out of the four creation and annihilation operators have acquired a factor of $i$, since only two of the original four $z(\bar{k},t)$, $x(\bar{k},t)$, $p_z(\bar{k},t)$, $p_x(\bar{k},t)$ operators have been transformed. In consequence, the sign of only one of the $[a_1(\bar{k}),\hat{a}_{1}(\bar{k}^{\prime})]$ and $ [a_2(\bar{k}),\hat{a}_{2}(\bar{k}^{\prime})]$ commutators is changed compared to (\ref{4.6}), and this eliminates the ghost signature in $[a_2(\bar{k}),a^{\dagger}_{2}(\bar{k}^{\prime})]$.

Now when a Hamiltonian is not Hermitian the action of it to the right and the action of it to the left are not related by Hermitian conjugation. Thus in general one must distinguish between right- and left-eigenstates, both for the vacuum and the states that can be excited out of it, and one must use the left-right inner product. This inner product obeys $\langle L(t)\vert R(t)\rangle=\langle L(0)\vert e^{iHt}e^{-iHt}\vert R(0)\rangle=\langle L(0)\vert R(0)\rangle$, to thus nicely be time independent in the non-Hermitian case. In the left-right basis we represent the equal-time $[y(\bar{k}),q(\bar{k^{\prime}})]=i$ and  $[x(\bar{k}),p_x(\bar{k}^{\prime})]=i$ commutators by  $q(\bar{k}^{\prime})=-i\overrightarrow{\partial_y(\bar{k}^{\prime})}$, $p_x(\bar{k}^{\prime})=-i\overrightarrow{\partial_x(\bar{k}^{\prime})}$ when acting to the right, and by $q(\bar{k})=i\overleftarrow{\partial_y(\bar{k}^{\prime})}$, $p_x(\bar{k}^{\prime})=i\overleftarrow{\partial_x(\bar{k}^{\prime})}$ when acting to the left. This then leads to right and left ground state wave functions of the form \cite{Bender2008b}
\begin{align}
\psi_0^R(y(\bar{k}),x(\bar{k}))&=\exp[-\tfrac{1}{2}(\omega_1+\omega_2)\omega_1\omega_2y^2(\bar{k})-\omega_1\omega_2y(\bar{k})x(\bar{k})-\tfrac{1}{2}(\omega_1+\omega_2)x^2(\bar{k})],
\nonumber\\
\psi_0^L(y(\bar{k}),x(\bar{k}))&=\exp[-\tfrac{1}{2}(\omega_1+\omega_2)\omega_1\omega_2y^2(\bar{k})+\omega_1\omega_2y(\bar{k})x(\bar{k})-\tfrac{1}{2}(\omega_1+\omega_2)x^2(\bar{k})].
\label{7.4}
\end{align}
Given these wave functions  the vacuum normalization for each $\bar{k}$ is given by \cite{Bender2008b}
 \begin{align}
 \langle \Omega^{L}(\bar{k})\vert \Omega^R(\bar{k})\rangle&=\int_{-\infty}^{\infty} dy(\bar{k})\int_{-\infty}^{\infty}dx(\bar{k})\langle \Omega^{L}\vert y(\bar{k}),x(\bar{k})\rangle\langle y(\bar{k}),x(\bar{k})\vert\Omega^R\rangle
 \nonumber\\
 &=\int_{-\infty}^{\infty} dy(\bar{k})\int_{-\infty}^{\infty}dx(\bar{k})\psi_0^L(y,x)\psi_0^R(y(\bar{k}),x(\bar{k}))
 =\frac{\pi}{(\omega_1\omega_2)^{1/2}(\omega_1+\omega_2)}.
 \label{7.5}
 \end{align}
Thus, just as we want, the left-right inner product normalization is finite. On normalizing to one we find that 
\begin{align}
\langle \Omega^{L}\vert \Omega^R\rangle&=\Pi_{\bar{k}} \langle \Omega^{L}(\bar{k})\vert \Omega^R(\bar{k})\rangle=
\Pi_{\bar{k}}1=1.
\label{7.17}
\end{align}
We thus confirm that the left-right vacuum normalization is both finite and positive. We thus establish the consistency and physical viability of the similarity-transformed higher-derivative scalar field theory. And we note that even though all the norms are  positive, the insertion of $\bar{\phi}$ into $-i\langle\Omega^{L}\vert T[\bar{\phi}(x)\bar{\phi}(0)]\vert \Omega^R \rangle$ (corresponding to  $+i\langle\Omega^{L}\vert T[\phi(x)\phi(0)]\vert \Omega^R \rangle$) generates the relative minus sign in $-[1/(k^2-M_1^2)-1/(k^2-M_2^2)]/(M_1^2-M_2^2)$ \cite{Bender2008b}. Thus with one similarity transform into an appropriate Stokes wedge we solve both the vacuum normalization problem and the negative-norm problem.

\section{The ultraviolet completion of Einstein gravity}
\label{S11}

As a quantum theory the standard second-order-derivative Einstein gravitational theory with its $1/k^2$ propagator is not renormalizable.  Since  graviton loops generate higher-derivative gravity terms, one can construct a candidate theory of quantum gravity by augmenting the Einstein Ricci scalar action with a term that is quadratic in the Ricci scalar. This gives a much studied \cite{footnote6a} quantum gravity action of the generic form
\begin{eqnarray}
I_{\rm GRAV}=\int d^4x(-g)^{1/2}\left[6M^2R^{\alpha}_{~\alpha}+(R^{\alpha}_{~\alpha})^2\right],
\label{11.1}
\end{eqnarray}
and it can be considered to be an ultraviolet completion of Einstein gravity. 
This same action also appears in Starobinsky's inflationary universe model \cite{Starobinsky1979}.

On adding on a matter source with energy-momentum tensor $T_{\mu\nu}$, variation  of this action with respect to the metric generates a gravitational equation of motion of the form
\begin{eqnarray}
-6M^2G^{\mu\nu}+V^{\mu\nu}=-\frac{1}{2}T^{\mu\nu}.
\label{11.2}
\end{eqnarray}
Here $G_{\mu\nu}$ is the Einstein tensor and $V_{\mu\nu}$ may for instance be found in \cite{Mannheim2006,Mannheim2017}, with these  terms being of the form 
\begin{align}
G^{\mu\nu}&=R^{\mu\nu}-\frac{1}{2}g^{\mu\nu}g^{\alpha\beta}R_{\alpha\beta},\quad
V^{\mu \nu}=
2g^{\mu\nu}\nabla_{\beta}\nabla^{\beta}R^{\alpha}_{~\alpha}                                             
-2\nabla^{\nu}\nabla^{\mu}R^{\alpha}_{~\alpha}                          
-2 R^{\alpha}_{\phantom{\alpha}\alpha}R^{\mu\nu}                              
+\frac{1}{2}g^{\mu\nu}(R^{\alpha}_{~\alpha})^2.
\label{11.3}
\end{align}                                 
If we now linearize about  flat spacetime with background metric $\eta_{\mu\nu}$ and fluctuation  metric $g_{\mu\nu}=\eta_{\mu\nu}+h_{\mu\nu}$, to first perturbative order we obtain 
\begin{align}
\delta G_{\mu\nu}&=\frac{1}{2}\left(\partial_{\alpha}\partial^{\alpha}h_{\mu\nu}-\partial_{\mu}\partial^{\alpha}h_{\alpha\nu}-\partial_{\nu}\partial^{\alpha}h_{\alpha\mu}+\partial_{\mu}\partial_{\nu}h\right)-\frac{1}{2}\eta_{\mu\nu}\left(\partial_{\alpha}\partial^{\alpha}h-\partial^{\alpha}\partial^{\beta}h_{\alpha\beta}\right),
\nonumber\\
\delta V_{\mu\nu}&=[2\eta_{\mu\nu}\partial_{\alpha}\partial^{\alpha} -2\partial_{\mu}\partial_{\nu}]
[\partial_{\beta}\partial^{\beta}h-\partial_{\lambda}\partial_{\kappa}h^{\lambda\kappa}],
\label{11.4}
\end{align}                                 
where $h=\eta^{\mu\nu}h_{\mu\nu}$. On taking the trace of the fluctuation equation around a flat background we obtain 
\begin{eqnarray}
[M^2+\partial_{\beta}\partial^{\beta}]\left(\partial_{\lambda}\partial^{\lambda}h-\partial_{\kappa}\partial_{\lambda}h^{\kappa\lambda}\right)=-\frac{1}{12}\eta^{\mu\nu}\delta T_{\mu\nu}.
\label{11.5}
\end{eqnarray}
In the convenient transverse gauge where $\partial_{\mu}h^{\mu\nu}=0$, the propagator for $h$ is given by
\begin{eqnarray}
D(h,k^2)=-\frac{1}{k^2(k^2-M^2)}=\frac{1}{M^2}\left(\frac{1}{k^2}-\frac{1}{k^2-M^2}\right).
\label{11.6}
\end{eqnarray}
As we see, in this case the $1/k^2$ graviton propagator for $h$ that would be associated with the Einstein tensor $\delta G_{\mu\nu}$ alone  is replaced by a $D(h,k^2)=[1/k^2-1/(k^2-M^2)]/M^2$ propagator. And now the leading behavior at large momenta is $-1/k^4$. In consequence, the theory is  thought to be renormalizable \cite{Stelle1977}. But since  $\langle \Omega\vert\Omega \rangle$ is not finite the proof of renormalizability has a flaw in it. Fortunately, the flaw is not fatal, and we rectify it below.

We recognize $D(h,k^2)$ as being of the same form as the scalar field propagator that was given in (\ref{4.2}), with $\phi$ being replaced by $h$ and with $M_1^2=M^2$, $M_2^2=0$. We can thus give $h$ an equivalent effective action of the form
\begin{eqnarray}
I_h&=&\frac{1}{2}\int d^4x\bigg{[}\partial_{\mu}\partial_{\nu}h\partial^{\mu}
\partial^{\nu}h-M^2\partial_{\mu}h\partial^{\mu}h\bigg{]}.
\label{11.7}
\end{eqnarray}
$I_h$  thus shares the same vacuum state normalization and negative norm challenges as the scalar  action given in (\ref{4.1}).

Thus if, as is conventional, we take $h$ to be Hermitian we would immediately encounter the negative-norm problem associated with the relative minus sign in (\ref{11.6}). However, since $M$ is Planck scale in magnitude, this difficulty can be postponed until observations can reach that energy scale. However, the lack of normalizabilty of the vacuum state has consequences at all energies and cannot be postponed at all. Specifically, with $\langle \Omega\vert\Omega \rangle$ being infinite we cannot even identify the propagator as $i\langle \Omega\vert T[h(x)h(0)]\vert \Omega \rangle$ since in analog to (\ref{1.8}) it will obey
\begin{align}
&(\partial_t^2-\bar{\nabla}^2)(\partial_t^2-\bar{\nabla}^2+M^2)D(h,x)=-\langle \Omega\vert\Omega \rangle\delta^4(x).
\label{11.8}
\end{align}
Consequently, we cannot make the standard Wick contraction expansion. And thus both the Feynman rules that are used presupposing that $\langle \Omega\vert\Omega \rangle$ is finite, and the renormalizability that is thought to then follow from them are therefore not valid. Additionally, with $\langle \Omega\vert\Omega \rangle$ being infinite, we cannot treat the Einstein theory with its $1/k^2$ propagator as an effective field theory that holds for momenta that obey $k^2 \ll M^2$.

However,  as noted above,  we can resolve all of these concerns by dropping the requirement that $h$ be Hermitian, and transform it to $i\bar{h}$, where $\bar{h}$ is self-adjoint in its own eigenstate basis. Then, with the theory being recognized as a $PT$ theory, vacuum state normalization and negative-norm problems are resolved and the theory is consistent. Moreover,  the propagator is given by $-i\langle \Omega^{L}\vert T[\bar{h}(x)\bar{h}(0)]\vert \Omega^R \rangle$ (corresponding to $+i\langle \Omega^{L}\vert T[h(x)h(0)]\vert \Omega^R \rangle$). And with the propagator still being given by (\ref{11.6}) as it satisfies $(\partial_t^2-\bar{\nabla}^2+M^2)(\partial_t^2-\bar{\nabla}^2)[-i\langle \Omega^{L}\vert T[\bar{h}(x)\bar{h}(0)]\vert \Omega^R \rangle]=-\delta^4(x)$, all the steps needed to prove renormalizability are now valid.  Consequently, the theory can now be offered as a fully consistent, unitary and renormalizable theory of quantum gravity  \cite{footnote6}.


\begin{thebibliography}{99}

\bibitem{Mannheim2022} P. D. Mannheim, \href{https://doi.org/10.48550/arXiv.2209.15047}{arXiv:2209.15047.} To appear in a special issue of Eur. Phys. J. Plus on \textit{Higher Derivatives in Quantum Gravity: Theory, Tests, Phenomenology}.


\bibitem{Mannheim2023} P. D. Mannheim, \href{https://arxiv.org/abs/2301.13029}{arXiv:2301.13029}.

 \bibitem{Bjorken1965} J. D. Bjorken and S. D. Drell, Relativistic Quantum Fields. McGraw-Hill, New York (1965).


\bibitem{Bender2008b} C. M. Bender and P. D. Mannheim, \href{https://doi.org/10.1103/PhysRevD.78.025022}{  Phys. Rev. D  \textbf{78}, 025022 (2008).}

\bibitem{footnote1} The difference in sign between the fourth-order $D(x)=i\langle \Omega\vert T[\phi(x)\phi(0)]\vert\Omega\rangle$ and the second-order $D(x)=-i\langle \Omega\vert T[\phi(x)\phi(0)]\vert\Omega\rangle$ is due to the fact that $[\phi,\dddot{\phi}]$ and $[\phi,\dot{\phi}]$ as respectively given in  (\ref{4.3}) and (\ref{1.2}) have opposite signs.

\bibitem{Pais1950} A. Pais and G. E. Uhlenbeck, \href{https://doi.org/10.1103/PhysRev.79.145}{Phys. Rev. \textbf{ 79}, 145 (1950).}


\bibitem{Mannheim2000} P. D. Mannheim and A. Davidson, \href{https://doi.org/10.48550/arXiv.hep-th/0001115}{arXiv:hep-th/0001115.}

\bibitem{Mannheim2005} P. D. Mannheim and A. Davidson, \href{https://doi.org/10.1103/PhysRevA.71.042110}{Phys. Rev. A \textbf{71}, 042110 (2005).}

\bibitem{Mannheim2007} P. D. Mannheim, \href{https://doi.org/10.1007/s10701-007-9119-7}{Found. Phys. \textbf{37}, 532 (2007).}

\bibitem{Bender2008a} C. M. Bender and P. D. Mannheim, \href{https://doi.org/10.1103/PhysRevLett.100.110402}{Phys. Rev. Lett. \textbf{100}, 110402 (2008).}

 \bibitem{mannheim2018antilinearity} P. D. Mannheim, \href{http://iopscience.iop.org/article/10.1088/1751-8121/aac035/meta}
{J. Phys. A: Math. Theor. \textbf{51}, 315302 (2018).}

 \bibitem{footnote2a} This path integral is also of interest for a separate reason. When evaluated with a real measure the Euclidean time path integral is well behaved [S. W. Hawking and T. Hertog, \href{https://doi.org/10.1103/PhysRevD.65.103515}{Phys. Rev. D \textbf{65}, 103515 (2002)}]. Since the Minkowski time path integral with a real measure is not, it means that  the contribution of the Wick contour in the Wick rotation from Minkowski time to Euclidean time cannot be neglected. In fact,  not only is it not zero, it must  even be infinite. Moreover, after the continuation into the complex plane that makes the Minkowski time path integral convergent, the same transformation makes the Euclidean time path integral divergent \cite{Mannheim2023}. So again the contribution of the Wick contour cannot be neglected. Since the Euclidean time path integral does not correctly capture the Minkowski time physics in this particular case, it would suggest that one would need to check whether or not this might be the case in any Euclidean field theory study.

 \bibitem{bender2010PT}  C. M. Bender and P. D. Mannheim, \href{https://doi.org/10.1016/j.physleta.2010.02.032}{Phys. Lett. A \textbf{374}, 1616 (2010).} 




\bibitem{footnote2}  If $AHA^{-1}=H$, where $A$ is a general antilinear operator and  $H\vert \psi\rangle=E\vert \psi\rangle$, then 
 $AH\vert \psi\rangle=AE\vert \psi\rangle=E^*A\vert \psi\rangle=AHA^{-1}A\vert \psi\rangle=HA\vert \psi\rangle$. Thus for every eigenvalue $E$ there is an eigenvalue $E^*$, with $H_{\rm PU}$ and $H_{S}$  being  in the $E=E^*$ realization in which  all energies are real. While this analysis would apply to any antilinear symmetry, since it was first discovered in the $PT$ context, theories with any antilinear symmetry are referred to  as  $PT$ theories. As discussed in \cite{mannheim2018antilinearity}, under the requirements solely of complex Lorentz invariance and  probability conservation one can show that a theory must be $CPT$ invariant, where $C$ denotes charge conjugation. When $C$ is separately conserved, $CPT$ defaults to $PT$. 


  \bibitem{bender2007making} C. M. Bender,  \href{https://doi.org/10.1088/0034-4885/70/6/R03}{Rep. Prog. Phys. \textbf{70}, 947 (2007).} 

\bibitem{bender2019pt} C. M. Bender, $PT$ Symmetry in Quantum And Classical Physics. World Scientific, Singapore (2019).









 \bibitem{footnote5a} While not being Hermitian $\bar{H}_S$ is self-adjoint when acting on its own eigenstates even though $H_S$ is not when it acts on its own eigenstates.  The original $H_S$ is composed of operators all of which are self-adjoint when they act on their own eigenstates, while not being self-adjoint when they act on the eigenstates of $H_S$. The continuation into the complex plane enables us to find an eigenstate basis within which both $\bar{H}_S$ and all the operators in it are self-adjoint. Since self-adjointness is requirement on boundary conditions, this is not affected by whether or not $\bar{H}_S$ might contain any factors of $i$. To be physically viable all that is needed of $\bar{H}_S$  is that its eigenvalues be real and that its left- and right-eigenstates can form an inner product $\langle L\vert R \rangle$ that is finite, positive and time independent. Its $PT$ symmetry ensures that this is the case.
 

 
  
 \bibitem{footnote6a} See e.g. the special issue referenced in \cite{Mannheim2022} and  the special issue of Il Nuovo Cimento C  \textbf {45}, Issue 2, March-April 2022 of contributions to the workshop on \textit{Quantum Gravity, Higher Derivatives and Nonlocality}. Included in this collection is P. D. Mannheim, \href{https://doi.org/10.1393/ncc/i2022-22027-6}{Nouvo Cim. \textbf{45}, 27 (2022)}, in which some of the background material presented in this paper may be found.
 
 \bibitem{Starobinsky1979} A. A. Starobinsky, \href{https://ui.adsabs.harvard.edu/abs/1979JETPL..30..682S/abstract} {JETP Lett. \textbf{30},  682 (1979).}

 
 \bibitem{Mannheim2006} P. D. Mannheim,   \href{https://doi.org/10.1016/j.ppnp.2005.08.001}{Prog. Part. Nucl. Phys.  \textbf{56}, 340 (2006).} 

 \bibitem{Mannheim2017} P. D. Mannheim, \href{https://doi.org/10.1016/j.ppnp.2017.02.001} { Prog. Part. Nucl. Phys. \textbf{94}, 125 (2017).} 
 
 \bibitem{Stelle1977} K. S. Stelle, \href{https://doi.org/10.1103/PhysRevD.16.953}{Phys. Rev. D \textbf{16}, 953 (1977};
  \href{https://doi.org/10.1007/BF00760427}{Gen. Rel. Gravit. \textbf{9}, 353 (1978).}







\bibitem{footnote6} Even though the $M^2$ field  now has positive norm, it still remains in the spectrum and would eventually have to be observed. As seen from (\ref{11.7}), the only reason that there is an $M^2$ term at all is because we are considering an action that has both second-order  and fourth-order terms. With a pure fourth-order theory there would be no dimensionful parameter in the action and the theory would be scale invariant. If like the gauge theories of $SU(3)\times SU(2)\times U(1)$ this scale symmetry is also local, we would be led to conformal gravity, a metric theory of gravity in which the action is left invariant under local changes of the metric of the form $g_{\mu\nu}(x)\rightarrow e^{2\alpha(x)}g_{\mu\nu}(x)$, where $\alpha(x)$ is a local function of the coordinates. The conformal gravity theory has been advocated and explored in \cite {Mannheim2006,Mannheim2017} and references therein. And 't Hooft has also argued [G.  't Hooft, \href{https://doi.org/10.1142/S0218271815430014}{Int. J. Mod. Phys. D   \textbf{24}, 1543001 (2015)}] that there should be an underlying local conformal symmetry in nature. In the conformal gravity theory the action is of the form $I_{\rm W}=-2\alpha_g\int d^4x(-g)^{1/2}\left[R_{\mu\kappa}R^{\mu\kappa}-(1/3) (R^{\alpha}_{~\alpha})^2\right]$, where $\alpha_g$ is a dimensionless  gravitational coupling constant The perturbative propagator has a $-1/k^4$ behavior at all $k^2$, and with its large $k^2$ behavior the theory is renormalizable [E. S. Fradkin and A. A. Tseytlin, \href{https://doi.org/10.1016/0370-1573(85)90138-3} {Phys. Rep. \textbf{119}, 233 (1985)}]. With a $-1/k^4$  propagator it would initially appear that there would be two massless particles at $k^2=0$. However, we cannot use the partial fraction decomposition given in (\ref{11.6}) as a guide since its $1/M^2$ prefactor is singular in the $M^2\rightarrow 0$ limit. Because of this singular behavior the $M^2=0$ Hamiltonian becomes of nondiagonalizable Jordan-block form and only has one massless eigenstate, with the other would-be massless eigenstate becoming nonstationary \cite{Bender2008b}.  With the theory being Jordan block there are now zero norm states. As well as the quantum gravity problem, conformal gravity also addresses the dark matter and dark energy problems. It has been shown (see \cite{Mannheim2006,Mannheim2017} and references therein) able to account for the systematics of galactic rotation curves without the need for any dark matter or its two free parameters per galactic dark matter halo, and able to account for the accelerating universe data without fine tuning.  It is a theory in which local and global physics are connected, with it recently having been shown that there is an imprint of galactic rotation curves on the recombination era cosmic microwave background (P. D. Mannheim, \href{https://doi.org/10.48550/arXiv.2212.13942}{arXiv:2212.13942}, Phys. Lett. B (in press)).




\end{thebibliography}
\end{document}